\documentclass[10pt,letterpaper,onecolumn,oneside]{article}
\usepackage[top=0.85in,left=1.75in,footskip=0.75in,marginparwidth=2in]{geometry}

\usepackage{amsmath}
\usepackage{booktabs} 
\usepackage{nopageno}
\newcommand{\bs}{\boldsymbol}

\usepackage[utf8]{inputenc}

\usepackage[colorlinks]{hyperref}

\usepackage[right]{lineno}

\usepackage{microtype}
\DisableLigatures[f]{encoding = *, family = * }

\usepackage{changepage}

\usepackage[aboveskip=1pt,labelfont=bf,labelsep=period,singlelinecheck=off]{caption}

\makeatletter
\renewcommand{\@biblabel}[1]{\quad#1.}
\makeatother

\usepackage{lastpage,fancyhdr,graphicx}
\usepackage{epstopdf}
\usepackage{color}

\definecolor{Gray}{gray}{.25}

\usepackage{graphicx}

\usepackage{sidecap}

\usepackage{wrapfig}
\usepackage[pscoord]{eso-pic}
\usepackage[fulladjust]{marginnote}
\reversemarginpar

\usepackage{comment}
\usepackage{siunitx}
\usepackage{textcomp}
\usepackage{units}
\usepackage{ulem}

\usepackage[round, numbers, sort&compress]{natbib}

\DeclareSIUnit\Molar{\textsc{m}} 

\begin{document}
\thispagestyle{plain}

\vspace*{0.35in}

\begin{flushleft}
{\Large
\textbf\newline{Traces of surfactants can severely limit the drag reduction of superhydrophobic surfaces}
}
\newline
\\
{Fran\c cois J. Peaudecerf}\textsuperscript{a},
{Julien R. Landel}\textsuperscript{b},
{Raymond E. Goldstein}\textsuperscript{a},
{and Paolo Luzzatto-Fegiz}\textsuperscript{c,1}
\\
\bigskip
\textsuperscript{a} Department of Applied Mathematics and Theoretical Physics, Center for Mathematical Sciences, University of Cambridge, Wilberforce Road, Cambridge CB3 0WA, UK
\\
\textsuperscript{b} School of Mathematics, University of Manchester, Manchester M13 9PL, UK
\\
\textsuperscript{c} Department of Mechanical Engineering, University of California, Santa Barbara, 93106, USA
\\
\bigskip
\textsuperscript{1} fegiz@engineering.ucsb.edu

\end{flushleft}

\section*{Abstract}
Superhydrophobic surfaces (SHSs) have the potential to achieve large drag reduction for internal and external flow applications. However, experiments have shown inconsistent results, with many studies reporting significantly reduced performance. Recently, it has been proposed that surfactants, ubiquitous in flow applications, could be responsible, by creating adverse Marangoni stresses. Yet, testing this hypothesis is challenging. Careful experiments with purified water show large interfacial stresses and, paradoxically, adding surfactants yields barely measurable drag increases. This suggests that other physical processes, such as thermal Marangoni stresses or interface deflection, could explain the lower performance.
To test the surfactant hypothesis, we perform the first numerical simulations of flows over a SHS inclusive of surfactant kinetics. These simulations reveal that surfactant-induced stresses are significant at extremely low concentrations, potentially yielding a no-slip boundary condition on the air--water interface (the "plastron") for surfactant amounts below typical environmental values. These stresses decrease as the streamwise distance between plastron stagnation points increases.
We perform microchannel experiments with thermally-controlled SHSs consisting of streamwise parallel gratings, which confirm this numerical prediction. We introduce a new, unsteady test of surfactant effects.
When we rapidly remove the driving pressure following a loading phase, a backflow develops at the plastron, which can only be explained by surfactant gradients formed in the loading phase. This demonstrates the significance of surfactants in deteriorating drag reduction, and thus the importance of including surfactant stresses in SHS models. Our time-dependent protocol can assess the impact of surfactants in SHS testing and guide future mitigating designs.

\vspace*{0.7in}

\section{Introduction}

Super-hydrophobic surfaces (SHSs) combine hydrophobic surface chemistry and micro- or nano-scale patterning in order to retain a network of air pockets when exposed to a liquid \citep[see e.g. reviews][]{Quere_ARMR_2008,Rothstein:2010im,Samaha12}. Since a large portion of the interface between the solid wall and the liquid is replaced by an air--liquid interface, which can be considered almost as a shear-free surface (known as a ``plastron''), SHSs could be used to obtain significant drag reduction in fluid flow applications \citep{Liu:2014cl,tian16}. Microchannel tests have recorded drag reductions of over 20\% \citep[e.g.][]{Ou_etal_PF_2004,Ou_Rothstein_PF_2005,Choi_Kim_PRL_2006, Joseph_etal_PRL_2006,tsai09,karatay13} and rheometer tests reported slip lengths of up to 185 $\mu$m \cite{Lee_Kim_PRL_2008}.
%
%
Turbulent flow experiments have reduced drag by up to 75\% \cite{Daniello_etal_PF_2009,woolford09,jung10,Park:2014bb}. However, a wide range of experiments have provided inconsistent results, with several studies reporting little or no drag reduction \cite{Park:2014bb,lee16,Watanabe_etal_JFM_1999,Gogte_etal_PF_2005,Henoch_etal_AIAA_2006,zhao07,peguero09,kim12,Bolognesi:2014ea,Gruncell_PhD_2014}.

\begin{figure}
\centering
\includegraphics[width=0.8\linewidth]{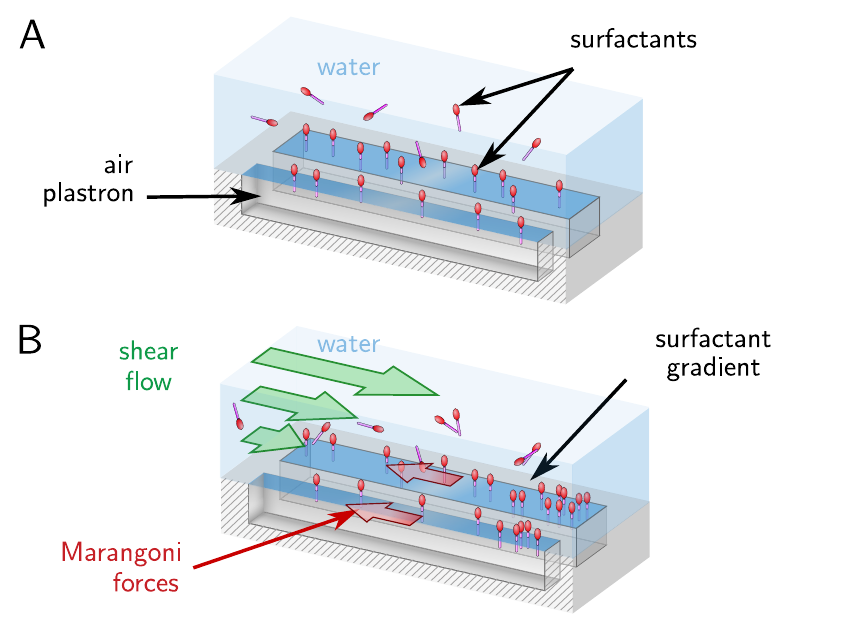}
\caption{Presence of surfactants as contaminant can generate Marangoni forces on a SHS. (\textit{A}) Surfactants present in water adsorb at the air--water interface of the plastron of SHSs. (\textit{B}) In the presence of an external flow, surfactants distribute in gradients between stagnation points. From these gradients result Marangoni forces resisting the flow and immobilizing the interface.}
\label{fig:Fig1}
\end{figure}

A key step towards solving this puzzle has come with the realization that surfactants could induce Marangoni stresses that impair drag reduction.
This was first hypothesized to account for experiments \cite{kim12,Bolognesi:2014ea} which revealed little measurable slip, in contradiction with available theoretical predictions based on the absence of surfactants \citep{Philip_ZAMP_1972a,Philip_ZAMP_1972b,cottinbizonne04,Lauga_Stone_JFM_2003,ybert07,davis09}. Following this hypothesis, surfactants naturally present in water would adsorb onto the air--water interface, as sketched in Fig.~\ref{fig:Fig1}\textit{A}, and they would be advected by the flow and therefore accumulate at downstream stagnation points, where the interface terminates in a three-phase contact line. The resulting surfactant gradient would then yield a Marangoni stress resisting the fluid motion, thereby decreasing slip and increasing drag (Fig.~\ref{fig:Fig1}\textit{B}). Traditional models of SHSs are surfactant-free, and therefore do not account for this additional drag. This is especially concerning for marine applications, since it is well documented that seawater contains significant amounts of surfactants \cite{Kropfli_etal_JGR_1999}. Rivers, estuaries, as well as fog also show significant levels of both synthetic and natural surfactants \cite{Ma1991,Facchini2000}.

Recent experiments have shown that while a nominally clean flow already displayed slip that was several times below predictions of surfactant-free theories, adding large amounts of surfactant had a barely measurable effect \cite{Schaffel:2016wd}. This counter-intuitive result appears to undermine the surfactant hypothesis, which would come with the expectation of a strong sensitivity to surfactant concentration. Other phenomena, such as thermal Marangoni effects (e.g. due to heating from a PIV laser), or interface curvature \citep{sbragaglia07,teo10}, must be assessed.  Proving or disproving the surfactant hypothesis, while resolving the above paradox,
 is essential to design SHSs that can achieve large, reliable drag reduction.

\section{Simulations with detailed kinetics}
To investigate the flow at extremely low surfactant concentrations, which are difficult to achieve and control in experiments, we developed a computational tool for simulating surfactant-laden flows over a SHS whose plane geometry is shown in Fig.~\ref{fig:simulations}\textit{A} (see \textit{Appendix} for details about the simulations). Neglecting air viscosity, we consider a two-dimensional air--water interface of length \(g\) on which surfactants from the bulk can adsorb/desorb and generate Marangoni forces. This interface (shown in pink in Fig.~\ref{fig:simulations}~\textit{A--C}) is on the top of a chamber of height \(H\) and bounded by no-slip ridges of length \(\ell/2\) each. The top of the chamber, of length $g+l$, represents a basis element of a SHS. The flow in the chamber is forced by a Poiseuille profile \(u(y)\) at $x=0$. 
In Fig.~\ref{fig:simulations}~\textit{B--D} we show simulation results from very low surfactant concentration to high concentration in the bulk, using the properties of the well-characterized surfactant sodium dodecyl sulfate (SDS) \cite{Prosser_etal_2001}. For very low concentration, $c_0=\SI{e-6}{\milli \Molar}$, the corresponding flow velocity is shown in Fig.~\ref{fig:simulations}\textit{B}, with gap length $g=\SI{100}{\micro\meter}$, wall length $\ell=\SI{50}{\micro\meter}$, channel height $H= \SI{100}{\micro\meter}$ and forced Poiseuille flow at $x=0$ of peak speed \SI{50}{\micro\meter\per\second}. The flow field is essentially identical to that for pure water, i.e. $c_0=0$.
Fig.~\ref{fig:simulations}\textit{D} shows the effect of the bulk surfactant concentration on the characteristic drag $\langle\tau\rangle$, which has been spatially averaged over one SHS unit of length $g+\ell$. The data are normalized by the corresponding wall shear stress $\tau_\mathrm{P}$ of the forcing Poiseuille flow.
The case $c_0=\SI{e-6}{mM}$ is highlighted by the left arrow in Fig.~\ref{fig:simulations}\textit{D}. Steady simulations with progressively larger bulk concentration, holding other parameters fixed, yield the curve shown in Fig.~\ref{fig:simulations}\textit{D}.
Fig.~\ref{fig:simulations}\textit{C} shows the flow velocity for $c_0=\SI{e-2}{mM}$. A large Marangoni stress appears and the slip velocity is reduced by more than one order of magnitude, practically reaching a no-slip boundary condition at the plastron. The corresponding drag is shown by the right arrow in Fig.~\ref{fig:simulations}\textit{D}.

For a bulk concentration of just $\SI{e-2}{\milli \Molar}$ (that is, \SI{e-5}{\mole\per\liter}, corresponding to less than 10 grams of surfactants per m$^3$ of water), the drag flattens to the no-slip asymptote, after which adding more surfactants has little effect. These results explain why adding surfactants in experiments has a barely measurable effect in most experiments, since the transition to a regime where Marangoni stresses dominate occurs at extremely small concentrations. Such small concentrations are very difficult to achieve in laboratory experiments, where controlling surfactant contamination (e.g. from SHS manufacturing and materials, water handling, micro-PIV beads, or any surface or fluid, including air, in contact with the water flowing through the microchannel) is exceedingly difficult. Needless to say, such levels of cleanliness are not found in flow applications. In addition, surfactants at such low concentrations are essentially impossible to detect using a classical tensiometer apparatus, since the corresponding decrease in surface tension, in a static fluid, is negligible \cite{Prosser_etal_2001}.

The results presented in Fig.~2\textit{D} are generic to other types of surfactant, as shown in \textit{Appendix} and Fig.~\ref{fig:figS2}. We also note that the properties of the SDS surfactant used in our study do not induce the strongest Marangoni stresses. In fact, its effect is rather mild compared to  the  surfactants reported in \citep{Chang_Franses_1995}. In normal flow applications, we should therefore assume that stronger surfactants than SDS are likely to be present, thus  deteriorating even  more  the performance of the SHS.

The inevitability of surfactants suggests that the main strategy to minimize Marangoni stresses is to optimize the SHS geometry. To test this hypothesis, we simulate flow over SHSs of varying interface length $g$. By increasing the distance $g$ between the upstream and downstream stagnation points, we reduce the average surfactant gradient over the plastron. Fig.~\ref{fig:simulations}\textit{E} shows drag versus interface length for $c_0=\SI{e-2}{mM}$ (plotted with red squares). We find that the drag over a contaminated surface is very sensitive to this change of geometry. The drag is also significantly larger than the idealized results for a perfectly clean flow (plotted with red diamonds) over a large range: $0.01 \leq g \leq \SI{10}{\milli\meter}$. 

\begin{figure}[t!]
\centering
\includegraphics[width=1\linewidth]{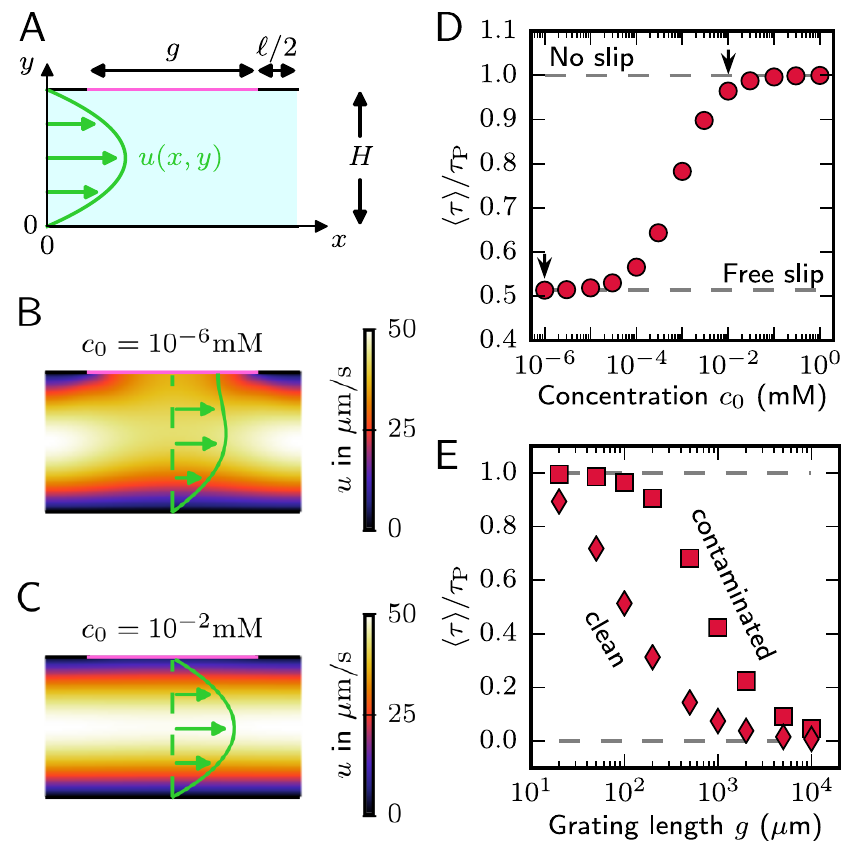}
\caption{Numerical simulations of surfactant-laden flows over a SHS. (\textit{A}) 2D geometry of the simulations for a periodic chamber (i.e. one SHS unit) with gratings of length $g = \SI{100}{\micro\meter}$, separated with ridges of length $\ell= \SI{50}{\micro\meter}$, and with forcing Poiseuille flow at $x=0$. (\textit{B}) Streamwise velocity $u$ for  bulk surfactant concentration $c_0 = \SI{e-6}{\milli\Molar} $, exhibiting slip on the plastron; (\textit{C}) at $c_0 = \SI{e-2}{\milli\Molar}$ no slip is reached on the plastron. (\textit{D}) Average normalised drag versus surfactant concentration. An asymptote is reached already at concentrations well below those found in the environment. (\textit{E}) Drag versus grating length, for perfectly clean water (red diamonds) and flow with bulk concentration of $c_0 = \SI{e-2}{\milli\Molar} $ (red squares), showing that drag reduction in surfactant-laden flows is very sensitive to grating length. All simulations are performed with a peak velocity of the forcing Poiseuille flow of \SI{50}{\micro\meter\per\second}.}
\label{fig:simulations}
\end{figure}


\section{Experiments show reduced slip} 

\begin{figure}[t!]
\centering
\includegraphics[width=1\linewidth]{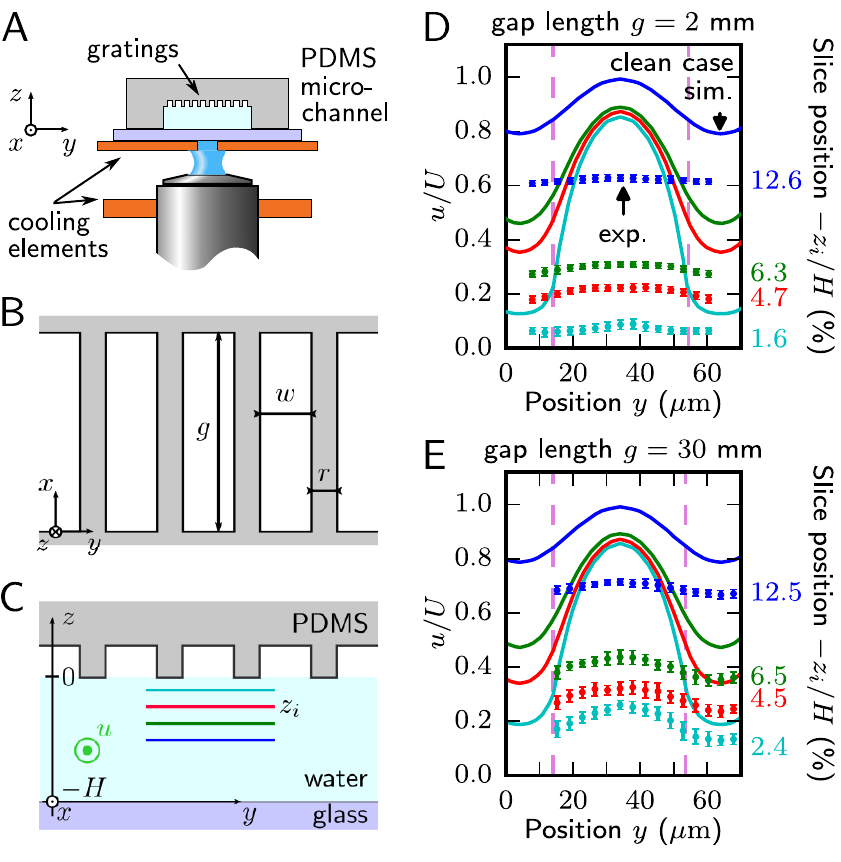}
\caption{$\mu$-PIV set-up and experimental velocity measurements with constant background flow. (\textit{A}) Side view of the inverted PDMS microchannel with a textured SHS on top. The microchannel is in contact with a thermal controller and lies above the water objective of the confocal microscope. (\textit{B}) Geometry of the streamwise parallel rectangular gratings forming the SHS. (\textit{C}) Cross-section in $(y,z)$ plane at $x=g/2$ of the (inverted) microchannel. The measuring planes of the $\mu$-PIV at heights $z_i$ are indicated with different colors. (\textit{D}) Lateral distribution of the streamwise velocity at different heights $z_i$ from the plastron, in the case of short lanes  $g=\SI{2}{mm}$. The profile is centered on a grating, with the edges of the ridges in dashed magenta lines. Experiments show no signicant slip at the plastron compared with clean case simulations. (\textit{E}) Same as in (\textit{D}) in the case of long lanes $g=\SI{30}{mm}$. Slip velocity at the plastron is still  three times smaller than predictions from clean case simulations.}
\label{fig:Fig3}
\end{figure}

To verify the effect of lane length on drag reduction, we performed experiments utilizing micro-particle image velocimetry ($\mu$-PIV) on a confocal microscope, and measured the velocity field of gravity-driven microchannel flows in planes parallel to a SHS (see Fig.~\ref{fig:Fig3}~\textit{A--C} and \textit{Appendix} for details). Similar to earlier work \citep{Daniello_etal_PF_2009,tsai09,Bolognesi:2014ea}, our SHSs were made of hydrophobic polydimethylsiloxane (PDMS) using soft photolithography techniques. The SHSs consisted of streamwise parallel rectangular lanes or gratings, as shown in Fig.~\ref{fig:Fig3}\textit{B}. The lane width $w=\SI{40}{\micro\meter}$ and ridge width $r=\SI{20}{\micro\meter}$ were kept fixed in the experiments. The microchannel height varied slightly, $100\leq H\leq \SI{120}{\micro\meter}$. We sealed the PDMS chamber using a glass coverslip held in place through Van der Waals adhesion, without surface treatment, to avoid surfactant contamination and to preserve PDMS hydrophobicity. The glass coverslip and chamber were maintained at a fixed temperature within $0.1^{\circ}\mathrm{C}$ accuracy through cooling elements. This minimized any potential thermal Marangoni effect (as discussed in \textit{Appendix}) and prevented significant condensation inside the gratings, thus ensuring stability of the plastron over several hours. We used  purified water in the experiments and cleaned all possible surfaces in contact with water and micro-beads following a strict cleaning protocol (\textit{Appendix}). The microchannel is connected to inlet and outlet reservoirs whose heights are adjusted separately to allow control of both the background pressure gradient and hydrostatic pressure in the microchannel. The plastron was maintained flat through adjustment of the hydrostatic pressure and the interface position was located with an accuracy of \SI{2}{\micro\meter}.

The symbols in Fig.~\ref{fig:Fig3}\textit{D} show experimental velocity profiles at several horizontal slices in the case of short lanes, $g=\SI{2}{mm}$ (standard deviations in the velocity are indicated with vertical bars). For comparison, Fig.~\ref{fig:Fig3}\textit{D} also shows surfactant-free numerical predictions (plotted with solid lines), for the same three-dimensional geometry. 
%
%
Similar to theoretical calculations assuming free slip at the plastron, the surfactant-free numerical simulations predict a very large slip, $\max(u/U) \approx 0.85$ at $z_i/H =1.6$\% from the plastron with $U$ the mean flow speed, whereas experimental data at $z/H\leq 2\%$ show no statistically significant slip. In the case of long lanes (Fig.~\ref{fig:Fig3}\textit{E}), $g=\SI{30}{mm}$, experimental data reveal some slip, which is still significantly lower than numerical and theoretical results \citep{Philip_ZAMP_1972a}, by approximately $70$\%. These results agree with our surfactant-laden simulations reported earlier in Fig.~\ref{fig:simulations}\textit{E}: for a given background flow and surfactant concentration, increasing the lane length increases the slip velocity, bringing it closer to predictions from surfactant-free models. This consistency  strongly suggests that the reduced slip velocity observed on our experimental SHSs with respect to theoretical predictions comes indeed from the presence of surfactants on the plastron surface.

\section{Pressure-relaxation experiments for surfactant effect} 

\begin{figure}[t!]
\centering
\includegraphics[width=1\linewidth]{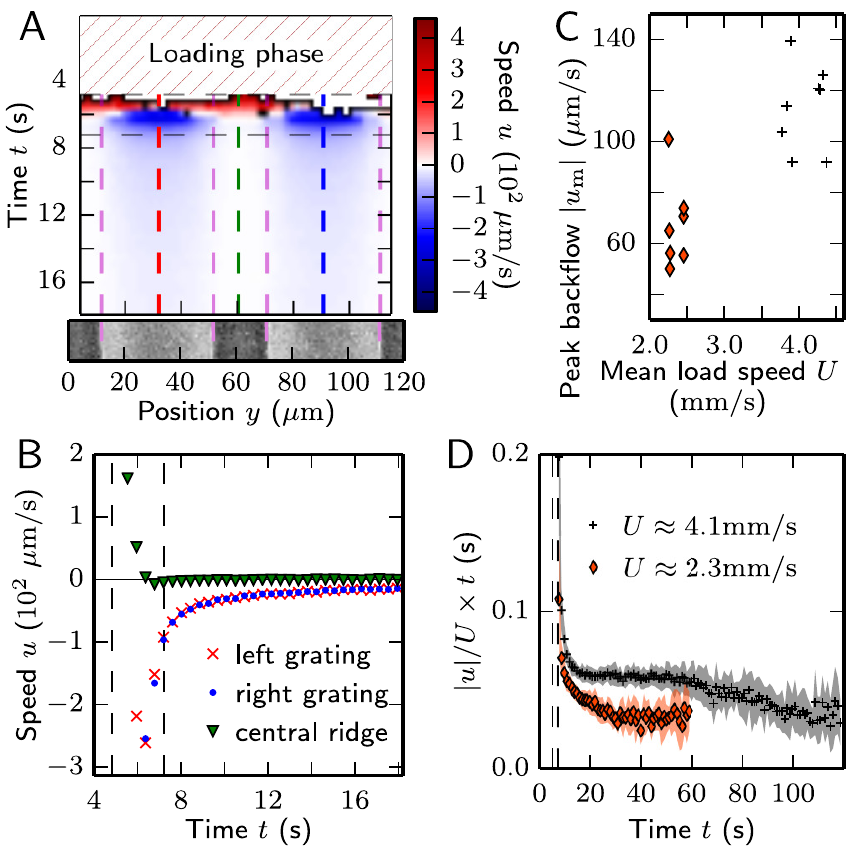}
\caption{Velocity measurements in pressure-relaxation experiments. (\textit{A}) Kymograph showing the time evolution of the lateral profile of the mid-gap streamwise velocity measured at $(x=g/2,z=\SI{-5}{\micro\meter})$ after a constant loading speed $U \approx \SI{4.1}{\milli\meter\per\second}$. The two horizontal black dashed lines designate the start and end of the pressure-head reduction to zero. Bottom inset is a picture showing the position of the two gratings studied (appearing in light grey). The kymograph clearly shows a strong backflow over the lanes but not over the ridge. (\textit{B}) Profiles from \textit{A} taken along the red, blue and green dashed vertical lines at the centerline of each grating and of the ridge, respectively. Red symbols are hidden by blue symbols as both gratings have the same profile. The standard deviation of all the data is smaller than the symbol size. (\textit{C}) Peak backflow velocity measured when the background pressure vanishes (corresponding to the second vertical dashed line in \textit{B}), versus the mean loading phase speed for the two sets of experiments conducted for $U\approx\SI{2.3}{\milli\meter\per\second}$ and $U\approx\SI{4.1}{\milli\meter\per\second}$. (\textit{D}) Ensemble average of the normalised backflow velocity multiplied by time for the two sets of experiments. Standard deviations are shown in lighter shading.} 
\label{fig:Fig4}
\end{figure}

To prove experimentally the surfactant hypothesis while circumventing the difficulty of removing significant traces of surfactants, we designed time-dependent pressure-relaxation experiments to reveal in-situ the presence of surfactant-induced stresses. All the experiments were conducted with the apparatus previously presented in Fig.~\ref{fig:Fig3}~\textit{A--C}, with controlled temperature and interface deflection. 
These experiments begin with a ``loading phase'', during which the flow is driven by a fixed, strong pressure gradient and allowed to reach steady state (see \textit{Appendix} for further details about the experimental protocol). The driving pressure is then rapidly decreased to zero while keeping the hydrostatic pressure approximately constant.
The change in background pressure-gradient is simply achieved by moving the stage onto which the inlet reservoir is attached to the exact level for which both inlet and outlet water levels are at the same vertical position. The motion of the stage was conducted very fast, within less than four seconds. During the loading phase, which was maintained for four minutes to ensure a steady state was reached, we measured a classical Poiseuille profile in the microchannel (Fig.~\ref{fig:figS1}), giving a mean loading speed $U$ (see \textit{Appendix} for further details on the technique used to measure $U$). Our measurements do not reveal any significant slip on the SHS side during the loading phase. However, a clear and somewhat unexpected backflow, with negative values (shown in blue in Fig.~\ref{fig:Fig4}\textit{A}), can be observed after the background pressure-gradient vanishes.

Fig.~\ref{fig:Fig4}\textit{A} shows the time evolution of the lateral profile of the streamwise velocity measured at $x=g/2$ and $z=\SI{-5}{\micro\meter}$ for a grating with $g = \SI{30}{mm}$. 
The background pressure-gradient decreases from a large constant value during the loading phase (mean loading speed in the microchannel $U \approx \SI{4.1}{\milli\meter\per\second}$ for $t\leq \SI{5}{\second}$) to zero at $t=\SI{7.2}{\second}$. 

As the flow in the microchannel is in the Stokes regime ($Re=HU_\mathrm{max}/\nu \approx 0.6$, with $\nu$ the kinematic viscosity of water), we expect that, in the absence of surfactants, the velocity should decrease from positive values to zero instantaneously and everywhere as soon as the background forcing vanishes. This monotonic behavior can be observed in Fig.~\ref{fig:Fig4}\textit{A} for the velocity measured above the ridge (see also green triangles in Fig.~\ref{fig:Fig4}\textit{B}). In contrast, the velocity above the plastron (plotted in red and blue in Fig.~\ref{fig:Fig4}\textit{B} for the middle of the left and right gratings, respectively) decreases sharply to large negative values during the stage motion, corresponding to a flow opposed to the background pressure gradient. This backflow persists for a long time after the background pressure gradient vanishes. In this experiment, it decays over approximately one minute. The same qualitative results and trends are obtained independently of the choice of grating, the direction of the loading flow and its intensity, as long as it is sufficiently strong.

During the loading phase, owing to large advection in the flow compared with diffusion and adsorption/desorption kinetics, surfactants at the interface are transported towards the downstream stagnation point of the lane. The concentration of surfactants increases significantly near this point, whereas it decreases everywhere else along the interface. This regime is likely to be analogous to the stagnant cap regime described for air bubbles rising in water \citep{PALAPARTHI:2006je}. When the background pressure gradient vanishes, the surfactant-induced  stresses are not opposed by viscous stresses any longer, and a  Marangoni backflow develops to homogenize the surfactant concentration at the plastron, as shown in Fig.~\ref{fig:Fig4}\textit{B}. A shear flow establishes across the height of the chamber, owing to the wall shear stress on the opposite side of the plastron ($z = - H$). As the distribution of surfactant becomes more uniform with time, surfactant-induced stresses decrease and the backflow velocity diminishes, as shown by the nonlinear temporal trend in Fig.~\ref{fig:Fig4}\textit{B}.

We find that the maximum velocity of the backflow $|u_\mathrm{m}|$, measured when the background pressure gradient vanishes ($t\approx \SI{7}{s}$), increases with the magnitude of the mean load speed $U$, as shown in Fig.~\ref{fig:Fig4}\textit{C}. This is consistent with a sharpening of the concentration gradient  near the downstream stagnation point at higher background pressure gradients during the loading phase, which then results in stronger Marangoni stresses driving the backflow. The observation that the plastron always displays a transient that reverses direction relating to the loading flow is a strong indication that Marangoni stresses are driven by surfactant accumulation, rather than by thermal gradients set up by laser or by other elements of the experimental apparatus.

To provide additional support to the surfactant hypothesis, as well as to show that the occurrence of the backflow is largely independent of the type of surfactant (and therefore of the associated kinetics), we develop a model for the backflow, and compare the resulting scaling for the plastron velocity to experiments. As surfactant diffusion is negligible at the air--water interface compared with advection, we model the backflow using a one-dimensional time-dependent advection equation for the interfacial surfactant concentration (this model is explained in detail in \textit{Appendix}). We assume adsorption/desorption fluxes are also negligible. The surfactant transport equation is coupled with a viscous-Marangoni stress balance. Using a similarity solution, we find that the magnitude of the backflow decreases in time such that $|u| \propto 1/t$, which is in agreement with the experimental results plotted in Fig.~\ref{fig:Fig4}\textit{D}. Fig.~\ref{fig:Fig4}\textit{D} shows the time evolution of the ensemble average (symbols) and standard deviations (light shadings) of the normalized backflow velocity $|u|/U$ multiplied by time. We show data for two sets of experiments conducted with $U\approx\SI{2.3}{\milli\meter\per\second}$ and $U\approx\SI{4.1}{\milli\meter\per\second}$ in the loading phase, plotted respectively in orange and black. The model is not valid at early time, $t\leq \SI{10}{s}$, due to the non-instantaneous decrease of the background pressure-gradient occurring between the two vertical dashed lines, owing to the finite time required for the stage motion (approximately 3 to 4 seconds). At late times $t\geq \SI{60}{s}$, the plateau observed for the large forcing experiments seems to end, as the data decrease again. This is consistent with the fact that the model, which assumes a semi-infinite lane, is not valid at late times when the effect of the opposite stagnation point (i.e. the upstream stagnation point during the loading phase) is felt. Moreover, we can notice in Fig.~\ref{fig:Fig4}\textit{D} that the characteristic time scale of the backflow, measured as the mean of $(|u/U|t)$ in the plateau region, increases with the forcing speed $U$. This indicates a nonlinear effect  probably due to changes in the surfactant distribution at $t=0$ for different $U$. 

\section{Outlook} 
An important conclusion is that quantitative models of superhydrophobic slip must also account for surfactant effects, or they can drastically overestimate drag reduction. This is consistent, for example, with established approaches for predicting the rise speed of small gas bubbles, which are known to be sensitive to surfactant-induced Marangoni stresses \cite{PALAPARTHI:2006je}. 
It would be valuable to build reduced-order models to enforce efficiently surfactant effects in Navier-Stokes solvers, obviating the need for a full solution of the complex coupled surfactant transport problem.

From a practical standpoint, our numerical and experimental findings point to the need to focus efforts on textures with large  distances between upstream and downstream stagnation points. This explains why several experimental works have reported that the most consistent performance was achieved with long gratings \cite{Park:2014bb,Truesdell_etal_PRL_2006}, whereas disorganized textures (which are attractive from a manufacturing standpoint) have been the least effective \cite{Gruncell_PhD_2014}. We can also note that some of the largest slip lengths reported in the literature were achieved with annular gratings in a circular rheometer \cite{Lee_Kim_PRL_2008}. Annular gratings are effectively infinitely long lanes, without stagnation points, thus  preventing the formation of surfactant concentration gradient and adverse Marangoni stresses. However, applications of SHSs with annular geometries  are limited.

From a methodological standpoint, our experimental tests involving pressure relaxation also provide a new and simple way of measuring the magnitude of surfactant-induced effects, for a SHS immersed in a given liquid. The presence of a backflow immediately reveals a plastron whose dynamics are strongly susceptible to surfactants. This can be a useful test to guide the design of SHSs that are unimpaired by surfactant Marangoni stresses. Overall, we believe this work constitutes a significant advance in our fundamental understanding of superhydrophobic drag reduction, moving the field closer to SHSs that perform reliably in realistic applications.




\section{Materials and Methods}

\subsection{Numerical simulations}
The numerical simulations involve solving eight partial differential equations for the transport of mass, momentum and surfactant in the fluid interior and along the air--water interface \cite{PALAPARTHI:2006je} (see \textit{Appendix} for more details). The ridges and bottom walls of the chamber have a no-slip boundary condition. At the air--water interface we enforce continuity of the viscous stress with the Marangoni stress, and continuity of the bulk velocity with the interfacial velocity. Note that air drag is negligible here, and is not considered. Surfactants present in the interior layer near the plastron are adsorbed or desorbed following specific kinetics. We have chosen a well-characterized surfactant, namely sodium dodecyl sulfate (SDS), as our case study surfactant. The SDS properties are well described by Frumkin kinetics \cite{Prosser_etal_2001}. These equations are solved using a finite element method from COMSOL Multiphysics\textsuperscript{\textregistered} software, with a Poiseuille profile as a forcing inlet condition and a weak constraint added to a free-slip boundary condition in order to capture the Marangoni effects at the air--water interface. A refined mesh at the interface and near the endpoints, together with higher order elements of discretisation,  guarantee accuracy of the  numerical solutions (see \textit{Appendix} for more details).

\subsection{Microchannel preparation} Microchannels of PDMS (Sylgard\textsuperscript{\textregistered} 184) with a top surface consisting of long rectangular gratings were fabricated using soft photolithography \cite{McDonald2002} (see \textit{Appendix}). They were bonded to glass coverslips through natural PDMS adhesion. All glass slides and tubing were washed thoroughly with purified water before use in the experiments. The  cleaning protocol is described in \textit{Appendix}.

\subsection{Laser Scanning Confocal Microscopy} To measure the fluid velocity in the microchannel, we washed and suspended micro-beads (LifeTechnologies FluoSphere\textsuperscript{\textregistered} carboxylate \SI{0.5}{\micro\meter} diameter yellow/green 505/515) in  purified water (using Milli Q\textsuperscript{\textregistered} water purification system, EMD Millipore) and injected the resulting solution at $0.08\%$ solids in the chamber. Using a spinning disk confocal microscope for green fluorescence (Yokogawa CSU-X1 mounted on Zeiss Axio Observer Z1, with laser line \SI{488}{\nano\meter} with power at the objective level less than $\SI{155}{\micro\watt}$), we captured image sequences at distinct $z$-planes of the chamber on high sensitivity EMCCD (Photometrics Evolve 512 Delta), at a rate of approximately 10 frames per second. We used a water objective (Olympus LUMPLFLN 60XW) to guarantee a proper determination of the position of the imaging plane and allow deep imaging without loss of fluorescence brightness. A bright field picture of the ridges and gratings allowed to determine the position of the SHS with respect to the field of imaging. To prevent heating of the microchannel during imaging, it was placed in thermal contact with a heat conducting plate, itself in contact with Peltier elements driven by a thermoelectric controller (Laird Technologies MTTC1410). To absorb the heat from the objective, we mounted a second Peltier element around it, coupled to a similar thermoelectric controller to maintain its temperature at $23\pm 0.1^{\circ}\mathrm{C}$. Fans were also used to cool the hot plates of the Peltier elements.

\subsection{Flow control} 
To ensure the driving pressure has no fluctuations, the flow of water with fluorescent particles is driven by gravity. The microchannel is connected with flexible tubing (Tygon\textsuperscript{\textregistered} ND-100-80) to two reservoirs (polypropylene tube, Eppendorf), whose absolute and relative heights can be varied independently. By varying the absolute position of the two reservoirs while keeping their relative position such that no flow is observable in the chamber, we minimized the curvature of the air--water interface in the chamber by adjusting the hydrostatic pressure. The maximum interface deflection over the width of the interface was less than \SI{2}{\micro\meter}. By adjusting the relative position of the reservoirs, varying background pressure gradients could be imposed in the channel to generate flow. The variation of relative position was performed using a motorized stage (Thorlabs NRT150/M) driven by a precision controller (Thorlabs APT BSC201) guaranteeing a position precision of \SI{2}{\micro\meter} and rapid changes in pressure gradients (maximum acceleration: \SI{50}{\milli\meter\per\second\tothe{2}}, maximum speed: \SI{50}{\milli\meter\per\second}).

\subsection{Image analysis and micro-PIV} The images obtained by confocal microscopy were analyzed  with MATLAB\textsuperscript{\textregistered} using a customized version of the particle image velocimetry (PIV) toolbox mPIV \citep{MoriChang2003}. The correlation functions between adjacent frames were rescaled to correct for non uniform time steps between frames, and the resulting correlations functions were added over a pack of 10 to 30 frames before searching for peak correlations. This allowed to correct for Brownian motion uncertainties. For the time-relaxation experiments, the velocity profile in the chamber during the loading phase at high flow rate was obtained by measuring the mean length of the fluorescent particle streaks on the image and dividing by the exposure time.

\section{Acknowledgments}

We thank D. Page-Croft and the engineers of the G.K. Batchelor Laboratory for their technical support. We gratefully acknowledge financial support from the Raymond and Beverly Sackler Foundation, the Engineering and Physical Sciences Research Council, the European Research Council Grant 247333, Mines ParisTech, the Schlumberger Chair Fund, as well as from Churchill and Magdalene Colleges, Cambridge.

\nolinenumbers

\clearpage

\fancyhead[r]{}
\fancyhead[l]{}
\bibliographystyle{unsrt}
\bibliography{biblio}



\clearpage
\appendix





\fancyhead[r]{}
\fancyhead[l]{}

\section{Supporting Model}

\subsection{Equations for surfactant-laden flows}
We consider a two-dimensional microchannel flow over a finite-length superhydrophobic surface (SHS), with the  geometry presented in Fig.~2\textit{A}. Since we maintained a flat plastron in the experiments, and the present focus is on Marangoni stresses, here we assume that the air--water interface is flat. 

The continuity and Navier-Stokes equations for the flow of mass and momentum are coupled with the transport equation for a surfactant.
The two-dimensional velocity field is ${\boldsymbol{u}} = ({u},{v})$ for the $(x,y)$ directions, and the surfactant concentration field is ${c}$. 
The fluid has density $\rho$ as well as dynamic and kinematic viscosities $\mu$ and $\nu$. The surfactant has bulk diffusivity $D$. The transport equations are
\begin{eqnarray} 
\nabla \cdot \bs{u} &=& 0, \label{eq:divu}\\  
\frac{\partial \bs{u}}{\partial t} + \nabla \cdot (\bs{u}\bs{u})&=& -\frac{\nabla{p}}{\rho} + \nu \nabla^2 \bs{u}, \\  
\frac{\partial c}{\partial t} + \nabla \cdot (\bs{u}c)&=&  D \nabla^2 c. \label{eq:C} 
\end{eqnarray}
The inlet conditions consist of a Poiseuille flow with mean velocity $U$ and maximum velocity $u_\mathrm{max}$, and a specified bulk concentration $c = c_0$, whereas at the outlet we set 
\begin{eqnarray}
\frac{\partial u}{\partial x} &=& 0,\\
\frac{\partial c}{\partial x} &=& 0. 
\end{eqnarray}
On the solid surfaces at $y=0$ and $y=H$ 
\begin{eqnarray} 
\boldsymbol{u} &=& \boldsymbol{0},\\
\frac{\partial c}{\partial y} &=& 0.
\end{eqnarray}
At the plastron, an adsorption/desorption model is used to
couple the surfactant transport between the bulk and the interface
\begin{eqnarray}
D \left.\frac{\partial c}{\partial y}\right|_I  &=& -S(c_I,\Gamma), \label{eq:dCdy} \\
\frac{\partial \Gamma}{\partial t} + \frac{\partial  ({u_I}\,\Gamma)}{\partial x}&=& D_\mathrm{s} \frac{\partial^2 \Gamma}{\partial x^2}  + S(c_I,\Gamma), \label{eq:Gamma}
\end{eqnarray}
where $\Gamma$ is the interfacial concentration, $D_\mathrm{s}$ is the surface diffusivity, the subscript $I$ denotes quantities at the interface, and $S(c_I,\Gamma)$ encapsulates the adsorption model. In practice, the choice of adsorption kinetics turns out to be of weak importance, since surfactant effects are already very strong at extremely low surfactant concentrations (as shown in Fig.~2). For such low values of concentration, different kinetics models are essentially equivalent \citep{Chang_Franses_1995}. For definiteness, here we use Frumkin kinetics 
\begin{equation}
S(c_I,\Gamma) = \kappa_\mathrm{a} c_I (\Gamma_\mathrm{m}-\Gamma) -  \kappa_\mathrm{d}\, \Gamma \, \textrm{e}^{A\Gamma/\Gamma_\mathrm{m}} , \label{eq:S}
\end{equation}
where $\kappa_\mathrm{a}, \kappa_\mathrm{d}$ are the adsorption and desorption coefficients, ${\Gamma}_\mathrm{m}$ is the maximum packing interfacial concentration and $A$ is the interaction coefficient \citep{Prosser_etal_2001}.

On the air--water interface, the velocity field is coupled to the interfacial surfactant distribution through a balance between viscous and Marangoni stresses  \citep{Rosen_Kunjappu_Wiley_2012}
\begin{eqnarray}
v &=& {0}, \\
\mu \left. \frac{\partial {u}}{\partial y} \right|_I&=& -n {R}{T}  \left( \frac{\Gamma_\mathrm{m}}{\Gamma_\mathrm{m}-\Gamma} + A \frac{\Gamma}{\Gamma_\mathrm{m}} \right) \frac{\partial \Gamma}{\partial x}, \label{eq:MarangoniStress}
\end{eqnarray}
where $n$ is the surfactant style constant \citep{Chang_Franses_1995}, ${R}$ is the universal gas constant and ${T}$ is the absolute temperature. The boundary condition for $\Gamma$, where the interface meets each solid boundary at a stagnation point (i.e. upstream stagnation point: $(x=\ell/2,y=H)$, and downstream stagnation point: $(x=g+\ell/2,y=H)$, see Fig.~2\textit{A}), is given by
\begin{equation}
\frac{\partial \Gamma}{\partial x} = 0.  \label{eq:SurfNoFlux}
\end{equation}

\section{Supporting Materials and Methods}

\subsection{Surfactant-laden simulations}

The model presented above was implemented in COMSOL Multiphysics® in a two-dimensional finite element  numerical simulation. The geometry corresponding to Fig.~2\textit{A} was created using the values for the gap length $g$, the ridge length $\ell$, the chamber height $H$ and the maximum forcing speed of the Poiseuille flow $u_\mathrm{max}$  presented in Table~\ref{tab:simu}. All the other physical parameters of the simulations are also presented in Table~\ref{tab:simu}, and correspond to the well-characterized surfactant sodium dodecyl sulfate (SDS). The SDS properties are well described by Frumkin kinetics \citep{Prosser_etal_2001}.

When designing the mesh of the domain, we were  particularly careful about strong possible variations of some variables near the stagnation points at the beginning and end of the gap. The maximum size of elements near these points is equal to  \SI{0.01}{\micro\meter}. For all simulations with $g<\SI{1}{\milli\meter}$, the maximum element size on the interface is \SI{0.05}{\micro\meter}. For simulations with $g = \SI{1}{\milli\meter}$ and $g = \SI{2}{\milli\meter}$, the maximum element size on the interface is \SI{0.2}{\micro\meter}. Finally, for $g = \SI{5}{\milli\meter}$ and $g = \SI{10}{\milli\meter}$, a coarser mesh is used in the central part of the interface, \SI{1}{\milli\meter} away from the endpoints, with a maximum element size of \SI{2}{\micro\meter}. In the bulk, the maximum element size is \SI{10}{\micro\meter} for all simulations. 

To implement the model, we combine the Laminar Flow module with a Dilute Species Transport module of COMSOL for the transport equations in the bulk (\ref{eq:divu}--\ref{eq:C}). The equation for the transport of surfactant on the interface (\ref{eq:Gamma}) is implemented through a General Form Boundary PDE, with a source term corresponding to the kinetic flux $S$. This  flux also serves to implement the boundary condtion (\ref{eq:dCdy}) at the interface for the Dilute Species Transport module. The Marangoni forces resulting from the non-uniform distribution of surfactants at the interface modify the Laminar Flow, as stated in (\ref{eq:MarangoniStress}), through a weak contribution at the interface coupled to a free-slip boundary condition.

The flow in the simulated chamber is forced by an inlet velocity boundary condition corresponding to a Poiseuille velocity profile $u(y) = 4 u_\mathrm{max} y (H-y)/H^2$. The initial guess velocity profile for the stationary solver is set to this reference Poiseuille profile in the entire chamber.

In order to increase the accuracy of the computation, we  discretize the fluid flow with quadratic elements for the velocity field  and linear elements for the pressure field, quadratic elements for the concentration field in the bulk and the concentration field on the interface.

We use the MUMPS solver of COMSOL to solve for the steady-state of the system, with a relative tolerance of $10^{-5}$.

\begin{figure}[tb]
\centering
\includegraphics[width=.7\linewidth]{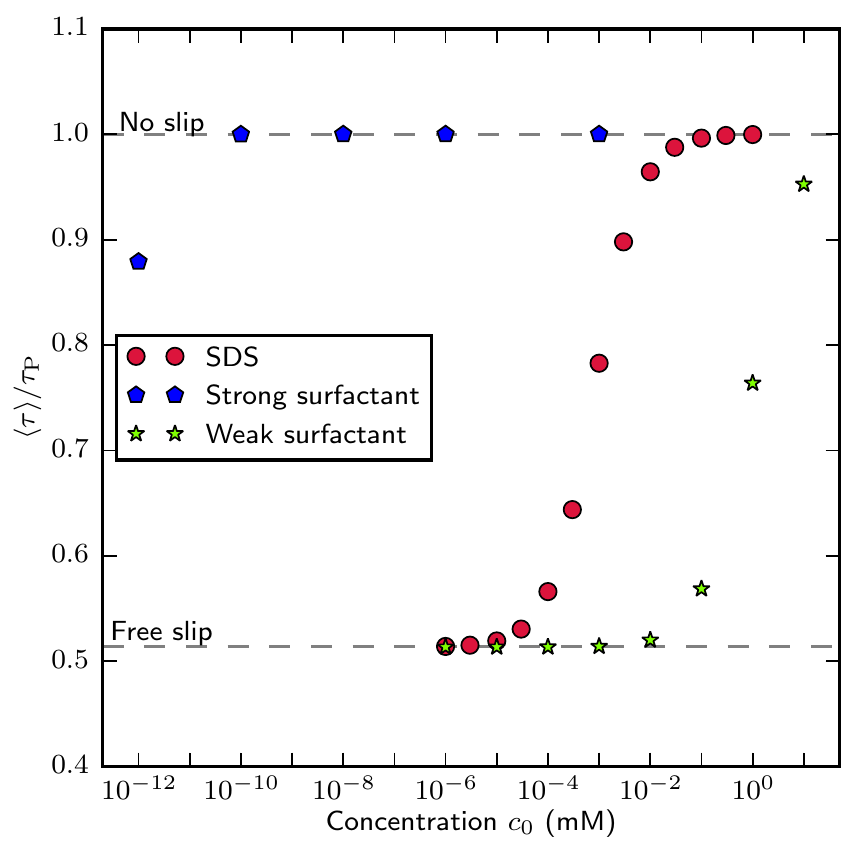}
\caption{Simulations of surfactant-laden flows in the model SHS chamber shown in Fig.~2\textit{A}.  We compare the average normalised drag versus the surfactant concentration for varying surfactant properties using Frumkin kinetics. The SDS properties are $\kappa_\mathrm{d} = \SI{500}{\per\second}$, $\kappa_\mathrm{a} = \SI{89.5}{\meter\tothe{3}\per\mole\per\second}$, $\Gamma_\mathrm{m} = \SI{3.92e-6}{\mole\meter\tothe{-2}}$, $A = -2.4$, $D = D_\mathrm{s} = \SI{7e-10}{\meter\tothe{2}\per\second}$. The model strong surfactant (defined using the extreme properties reported in \citep{Chang_Franses_1995}) has high affinity to the interface and low diffusivity: $\kappa_\mathrm{d} = \SI{1}{\per\second}$, $\kappa_\mathrm{a} = \SI{e6}{\meter\tothe{3}\per\mole\per\second}$, $\Gamma_\mathrm{m} = \SI{e-5}{\mole\meter\tothe{-2}}$, $A = -3$, $D = D_\mathrm{s} = \SI{e-11}{\meter\tothe{2}\per\second}$. Similarly, the model weak surfactant is defined with weak affinity to the interface and high diffusivity: $\kappa_\mathrm{d} = \SI{100}{\per\second}$, $\kappa_\mathrm{a} = \SI{e-1}{\meter\tothe{3}\per\mole\per\second}$, $\Gamma_\mathrm{m} = \SI{e-6}{\mole\meter\tothe{-2}}$, $A = 3$, $D = D_\mathrm{s} = \SI{e-9}{\meter\tothe{2}\per\second}$. The behaviour of the transition is similar for all three surfactants, but it occurs at different concentration thresholds. The threshold is particularly low for the strong surfactant.} 
\label{fig:figS2}
\end{figure}

In order to check how the results obtained would change for surfactants of different strengths, we also ran simulations for two extreme sets of parameters values for the surfactant choice, using the Frumkin kinetics framework. The first set corresponds to a model strong surfactant with high affinity to the interface and low diffusivity ($\kappa_\mathrm{d} = \SI{1}{\per\second}$, $\kappa_\mathrm{a} = \SI{e6}{\meter\tothe{3}\per\mole\per\second}$, $\Gamma_\mathrm{m} = \SI{e-5}{\mole\meter\tothe{-2}}$, $A = -3$, $D = D_\mathrm{s} = \SI{e-11}{\meter\tothe{2}\per\second}$), the second corresponds to a model weak surfactant with weak affinity to the interface and high diffusivity which promotes the smoothing of any interfacial gradients ($\kappa_\mathrm{d} = \SI{100}{\per\second}$, $\kappa_\mathrm{a} = \SI{e-1}{\meter\tothe{3}\per\mole\per\second}$, $\Gamma_\mathrm{m} = \SI{e-6}{\mole\meter\tothe{-2}}$, $A = 3$, $D = D_\mathrm{s} = \SI{e-9}{\meter\tothe{2}\per\second}$). These parameters were selected from the extreme values of data reported in Tables 1 and 3 in \citep{Chang_Franses_1995}. We performed simulations with the rest of the parameters as in the simulation for Fig.~2\textit{D}, except for the bulk concentration $c_0$ whose range was extended to cover the transitions for the strong and weak surfactant respectively. For both cases, we observe progressive immobilization of the interface and increase of viscous stress with increasing bulk concentration of surfactant (Fig.~\ref{fig:figS2}). For the strong surfactant, the transition towards a no-slip boundary condition occurs at minute concentrations, below $c_0 \approx \SI{e-12}{\milli \Molar}$, and well below the transition value for SDS ($c_0 \approx \SI{e-4}{\milli \Molar}$). For the weak surfactant, the transition occurs at $c_0 \approx \SI{1}{\milli \Molar}$, which is much higher than for SDS. In general, one cannot of course guarantee that no traces of strong surfactants are present in a given flow. 

\subsection{Surfactant-free simulations}
To obtain a reference flow profile for long rectangular gratings located on one side of a three-dimensional microchannel, in the idealized case of pure water ($c_0=0$), we solved the Navier-Stokes equation using COMSOL Multiphysics®. Our domain corresponds to the portion of a microchannel below half of a grating element, with plane of symmetry $(x,y=w/2,z)$ (see Fig.~2 \textit{B} and \textit{C}). We use symmetry boundary conditions on each side of the domain to solve for the flow on a large number of parallel gratings as in the experiment. The rectangular grating has a length $g = \SI{1}{\milli\meter}$, a width $w = \SI{40}{\micro\meter}$ and is bounded by lateral ridges of width $r = \SI{20}{\micro\meter}$. No-slip boundary conditions are imposed on the ridges, while free-slip is imposed on the plastron. The grating is preceded and followed by a no-slip area of length $\ell/2 = \SI{10}{\micro\meter}$. The channel height is $H = \SI{120}{\micro\meter}$. The flow was forced by a three-dimensional Poiseuille profile $u(z) = 4 u_\mathrm{max} z (H-z)$ with $u_\mathrm{max} = \SI{120}{\micro\meter\per\second}$. This same velocity profile was also chosen as initial guess for the steady-state solution. Water viscosity was $\mu = \SI{9.3e-4}{\newton\second\per\meter}$, corresponding to a temperature of water of \SI{23}{\degree}. We use a Physics-controlled mesh with Finer element size, with a linear discretization of elements. The results of this simulation were used for both plots of Fig.~3 \textit{D} and \textit{E}, as the velocity profile at the middle of the grating was found independent of the grating length in the simulation for $g \gg H$ and $g \gg w$.

\begin{table}[tb]
    \caption{Parameters for surfactant-laden simulations.}
    \centering
    \small
    \label{tab:simu}
    \sisetup{
        table-number-alignment = center,
        table-figures-integer = 1,
        table-figures-decimal = 1,
        table-figures-exponent = 1,
        table-sign-exponent,
        range-phrase = --
    }
    \begin{tabular}{l
                    c
                    S[table-number-alignment = right,table-alignment = right]
                    s[table-unit-alignment = left]
                    l
                    }
          \toprule
          {Parameter} & {Symbol} & {Value} & {Unit} & {Use} \\
          \midrule
          
          {Grating length} & $g$ & 1e2 & \micro\meter & {Fig.~2 \textit{B}--\textit{D}}\\
          {} & {} & {$20 \;\mathrm{to} \;100\times 10^2$} & \micro\meter & {Fig.~2\textit{E}}\\
          {Bulk concentration} & $c_0$ & 1e-2 & \mole\meter\tothe{-3} & Fig.~2\textit{D} \\
          {} & {} & {$10^{-6} \; \mathrm{to}\; 1$} & \mole\meter\tothe{-3} & Fig.~2 \textit{A}, \textit{B},  \textit{E} \\
          {} & {} & {$10^{-12} \; \mathrm{to}\; 10$} & \mole\meter\tothe{-3} & Fig.~\ref{fig:figS2} \\
          {Chamber height} & $H$ & 1e2 & \micro\meter & \\
          {Ridge length} & $\ell$ & 5e1 & \micro\meter & \\
          {Maximum forcing speed} & $u_\mathrm{max}$ & 5e1 & \micro\meter\per\second &  \\
          {Mean forcing speed} & $U$ & 3.3e1 & \micro\meter\per\second &  \\
          {Water viscosity} & $\mu$ & 8.9e-4 & \newton\second\meter\tothe{-2} &  \\
          {Water surface tension} & $\sigma_0$ & 72e-3 & \newton\per\meter &  \\
          {Bulk diffusivity} & $D$ & 0.7e-9 & \meter\tothe{2}\per\second &  \\
          {Surface diffusivity} & $D_\mathrm{s} $ & 0.7e-9 & \meter\tothe{2}\per\second &  \\
          {Desorption coefficient} & $\kappa_\mathrm{d} $ & 500 & \per\second &  \\
          {Adsorption coefficient} & $\kappa_\mathrm{a} $ & 89.5 & \meter\tothe{3}\per\second\per\mole &  \\
          {Max. packing concentration} & $\Gamma_\mathrm{m} $ & 39.2e-7 & \mole\meter\tothe{-2} &  \\
          {Interaction coefficient} & $A $ & -2.4 & - &  \\
          {Surfactant style constant} & $n $ & 2 & - &  \\
          \bottomrule
    
    \end{tabular}
\end{table}










\section{Supporting experimental protocols}

\subsection{Cleaning protocols}

Two different cleaning protocols were followed in the preparation of the experiments. For both cleaning protocols, as well as all the experiments conducted, only purified water (using Milli Q® water purification system, EMD Millipore) at \SI{23}{\degree C} with resistivity \SI{18.2}{\mega\ohm . \cm}$^{-1}$ and less than 5 parts per billion of total organic content was used. 

A  strict 10-day cleaning protocol was designed in an attempt to avoid  any  contamination of the microchannel, which could induce surfactant Marangoni stresses. The cleaning and experimental preparation  were performed using lab coats and thoroughly washed nitrile gloves (Fisherbrand) (we note that standard laboratory gloves have traces of chemicals on their surface which induce surfactant Marangoni stresses). The preparation of the polydimethylsiloxane (PDMS, Sylgard® 184) microchannels was done in a clean room. In this protocol, apart from the PDMS, only materials that could be cleaned thoroughly were used for the surfaces that were in contact with water during the experiments. In particular, all common plastic materials were avoided as they tend to release chemical traces with surfactant effect when in contact with water. All tubings were made of FEP (\SI{0.5}{\milli\meter} internal diameter, The Dolomite Centre Ltd), connectors were made of stainless steel, the syringes (Gastight Hamilton) used to handle water and the micro-bead suspension were made of glass and PTFE, and fitted with stainless steel needles (24 G injection needles, Carl Roth GmbH), the inlet and outlet reservoirs were made of glass. All the tubings, connectors, needles and reservoirs undergone five washing, rinsing, and curing cycles over a 10-day period prior to the experiments. The curing containers were large glass beakers that had been cleaned in an acid rinse dishwasher, and  further rinsed for 5 minutes with purified water.  During the curing process, all the beakers were covered to reduce contamination from the air. On the day of the experiments, all the tubings and reservoirs were washed again with purified water. Washed metal tweezers were used to handle tubings and connectors in order to avoid touching surfaces that could be in contact with the water flowing through the microchannel.
The fluorescent micro-beads (LifeTechnologies FluoSphere® carboxylate \SI{0.5}{\micro\meter} diameter yellow/green 505/515) used to perform $\mu$-PIV were washed and rinsed ten times with purified water to dilute significantly any potential surfactant contamination traces. The cover slip forming the base of the microchannel was washed with abundant purified water and then air dried.

This strict cleaning protocol was only followed when conducting some steady forcing experiments. The results, similar to those presented in  Fig.~3\textit{D} (conducted following the normal cleaning protocol described below), showed no or little slip in comparison to  theoretical and numerical predictions with surfactant-free flows.  As we show in Fig.~2\textit{D}, the level of contamination necessary to induce surfactant Marangoni stresses is extremely small, of the order of \SI{e-4}{\milli \Molar} for the SDS surfactant. Moreover, as we  show in Fig.~\ref{fig:figS2}, SDS is not the strongest surfactant and can be considered as inducing mild Marangoni stresses \citep[see Tables 1 and 3 in][for a comparison of a broad range of surfactants]{Chang_Franses_1995}. Therefore, it is very likely that, even  following this strict cleaning protocol, sufficient traces of chemicals with a surfactant effect contaminated our experiments. We believe that the most likely source of contamination in our experiments is the PDMS and its associated impurities. Uncrosslinked PDMS chains or impurities trapped in the PDMS could have a surfactant effect as has been observed in  \cite{hourlierfargette16}. Contamination could also come from other sources, which might simply be unavoidable in  normal laboratory conditions.

As surfactant contaminations were simply unavoidable in our experiments, a less time-consuming cleaning protocol was used for all the experimental results presented in this study. We used flexible tubing (Tygon® ND-100-80) instead of the FEP tubing, which were more difficult to handle due to their rigidity. The outlet and inlet reservoirs were replaced by polypropylene tubes (Eppendorf) or plastic syringes (BD Plastipack\texttrademark). All the other elements of the apparatus, preparation tools and materials were the same. Furthermore, cleaning of the tools, materials, tubings, connectors and reservoirs was performed on the day of the experiment. They were all washed with plenty of purified water: typically with at least ten times the volume they can contain. The fluorescent micro-beads were washed at least three times.

\subsection{Steady forcing experiments} 

For steady forcing experiments, the chamber was first filled with the suspension of microbeads, taking great care to avoid  trapping  any air bubble in the tubing or in the chamber. In order to obtain a low level of flow rate in the chamber while allowing accurate positioning of the inlet reservoir with respect to the outlet reservoir, a constriction was introduced on the hydraulic line by mounting a Gauge 30 polypropylene syringe tip (Adhesive Dispensing Ltd) on the inlet reservoir (syringe from BD Plastipack\texttrademark). Using a manual linear  \SI{150}{mm} motorized stage (Thorlabs NRT150/M) driven by a precision controller (Thorlabs APT BSC201, with \SI{2}{\micro\meter} precision,  maximum acceleration: \SI{50}{\milli\meter\per\second\tothe{2}}, maximum speed: \SI{50}{\milli\meter\per\second}), the inlet reservoir was moved vertically after initial filling of the chamber until no flow could be observed in the middle of the chamber. This corresponded to the level of zero pressure gradient along the microfluidic line. The inlet position was then shifted by $\SI{5}{\milli\meter} \pm \SI{10}{\micro\meter}$ using the linear manual stage.  Imaging was then performed in the central longitudinal portion of the gratings. Stacks of 30 successive images were taken at approximately 10\,fps at different $z$ positions in the chamber and at different time points for each experiment. Once a microchannel was successfully prepared to conduct a series of experiments, most of the plastrons of the SHS gratings remained stable for approximately two hours.

The details of the parameters for the experiments presented in Figure 3\textit{D-E} are described in Table~\ref{tab:expe}. The mean forcing speed $U$ was calculated from the experimental velocity profile by fitting a parabolic profile $u(z) = 4 u_\mathrm{max} z (H-z)$ to the data.

\begin{table}[tb]
    \caption{Parameters for steady-forcing experiments.}
    \centering
    \label{tab:expe}
    \sisetup{
        table-number-alignment = center,
        table-figures-integer = 1,
        table-figures-decimal = 1,
        table-figures-exponent = 1,
        table-sign-exponent,
        range-phrase = --
    }
    \begin{tabular}{l
                    c
                    S[table-number-alignment = right,table-alignment = right]
                    S[table-number-alignment = right,table-alignment = right]
                    s[table-unit-alignment = left]
                    }
          \toprule
          {Parameter} & {Symbol} & \multicolumn{2}{c}{Value} & {Unit} \\
          {} & {} & {Fig.~3\textit{D}} & {Fig.~3\textit{E}} & {} \\
          \midrule
          {Grating length} & $g$ & 2e3 & 3e4 & \micro\meter \\
          {Grating width} & $w$ & 4e1 &4e1 & \micro\meter \\
          {Chamber height} & $H$ & 1.3e2 &9.9e1 & \micro\meter  \\
          {Ridge width} & $r$ & 2e1 &2e1 & \micro\meter  \\
          {Max. forcing speed} & $u_\mathrm{max}$ & 1.1e2 & 1.3e2 &\micro\meter\per\second   \\
          {Mean forcing speed} & $U$ & 7.3e1 &8.7e1 & \micro\meter\per\second   \\
          {Temperature} & $T$ & 23&23 & \degreeCelsius  \\
          \bottomrule
    \end{tabular}
\end{table}

\subsection{Pressure-relaxation experiments}

\begin{figure}[t]
\centering
\includegraphics[width=0.75\linewidth]{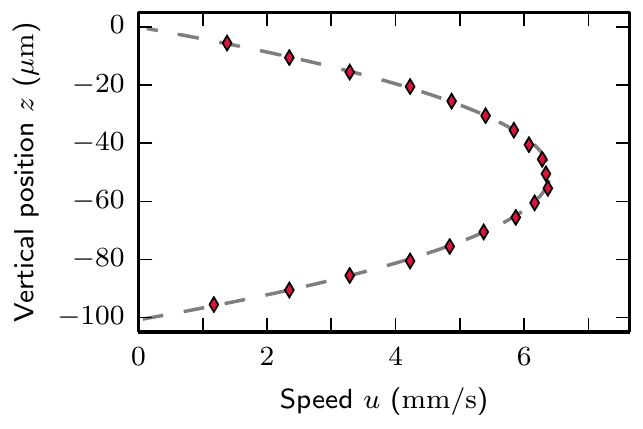}
\caption{Vertical distribution of the streamwise velocity profile during the loading phase of the experiment presented in Fig.~4\textit{A}. The dashed line shows the result of a parabolic fit of the velocity profile. The mean speed of the flow extracted from this fit is $U \approx \SI{3.9}{\milli\meter\per\second} $.} 
\label{fig:figS1}
\end{figure}

The protocol for each experiment had two phases: an initial loading phase with strong background flow and then a second phase without background flow to measure the surfactant Marangoni driven backflow. During the initial loading phase the flow  was driven at very high background pressure gradient, in order to transport any surfactant along the air--water interface to the downstream stagnation end of a grating. This phase lasts for four minutes, during which images of the flow field were taken at different heights in the channel in order to obtain the vertical distribution of the streamwise velocity profile. A typical  velocity profile measured during the loading phase of the experiment shown in Fig.~4\textit{A} is presented in Fig.~\ref{fig:figS1}. As the exposure time on the camera was smaller but of the same order of magnitude than the typical time for the fluorescent particles to cross the field of view, particles were seen as streaks of dots on each image. The dots forming the streaks originate from the imaging by the spinning disk used with the microscope. The velocity profile was obtained by measuring the mean length of the fluorescent particle streaks on the image and dividing by the exposure time. As we can see in  Fig.~\ref{fig:figS1}, the velocity profile measured in a vertical plane centered in the middle of a grating does not show any significant slip velocity at the air--water interface. We note that the spatial resolution and the technique to measure the velocity were not designed to measure accurately the velocity close to the plastron, as was done in the steady forcing experiments (see previous section and Fig.~3). The aim here was to obtain an estimate of the mean flow field in the microchannel. The profile follows a classical two-dimensional Poiseuille profile, from which we computed the mean loading speed $U$ plotted in Fig.~4\textit{C}.

\begin{figure}[t!]
\centering
\includegraphics[width=0.7\linewidth]{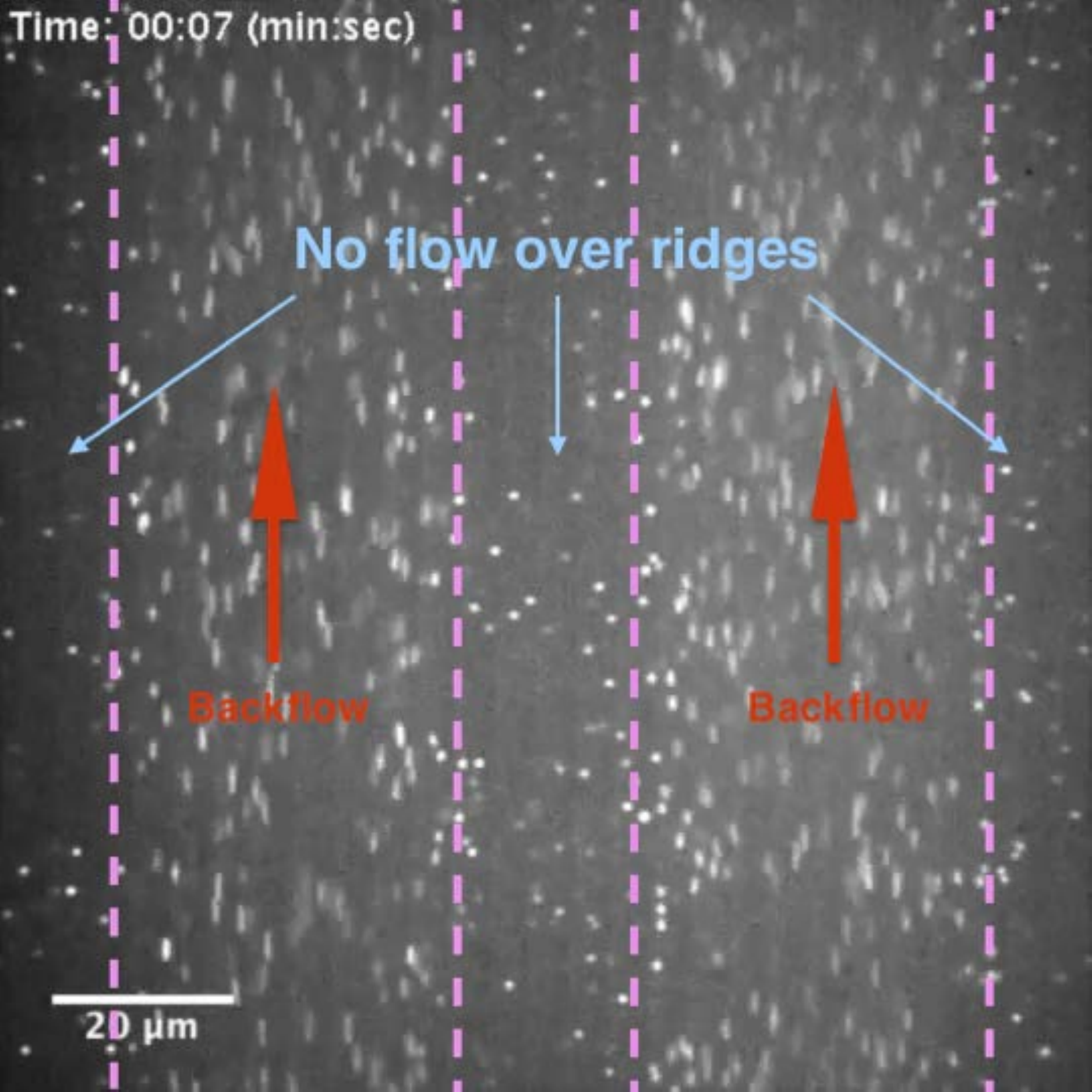}

\vspace{10pt}

\caption{The loading flow was downwards in this image, such that an upward motion corresponds to a backflow, as shown by the red arrows. This corresponds to Exp.~1-6 in Table~\ref{tab:unsteadyexp}.} 
\label{fig:MovieS1}
\end{figure}

In the second  phase, the background pressure gradient was suppressed to stop the flow. Images of the flow field were recorded at a distance of $z=\SI{-4}{\micro\meter} \pm \SI{1}{\micro\meter}$ from the air--water interface. Images were recorded for 1 or 2 minutes starting approximately five seconds before the background pressure gradient was suppressed (this corresponds to $t=0$ in Fig.~4 \textit{A}, \textit{B} and \textit{D}), at a frame rate of around 24 frames per second. 
An example is shown in Fig.~\ref{fig:MovieS1}.
These images were then analyzed with $\mu$-PIV, as explained in the Materials and Methods section, to produce the velocity fields, such as the one displayed in Fig.~4\textit{A}. Then, the experiment was repeated but with imposing a negative background pressure gradient in the microchannel in order to produce an opposite flow. We could thus verify that the effect observed was independent of the flow direction in the initial loading phase.

The apparatus and the microchannels were  prepared  following the initial steps of the protocol described in the previous section.
Then, to control the background pressure gradient imposed in the microchannel during the loading phase and ensure a rapid and smooth transition between the two phases, the inlet  reservoir was attached at mid-height onto the linear \SI{150}{mm} motorized stage. This mid-height constituted our zero elevation reference. The outlet reservoir was attached onto the fixed part of the motorized stage, which was itself attached onto a millimetric precision vertical ramp. Adjusting the height of the motorized stage effectively controlled the hydrostatic pressure in the microchannel and thus the interfacial deflection of the plastron. An initial upward loading phase was conducted, without recording any images, in order to determine accurately the level of zero background pressure gradient required for the second phase. Indeed, during this first loading phase where the inlet reservoir was raised to a given height ($\Delta H_r>0$) compared with the outlet reservoir, water  transferred from the inlet reservoir to the outlet reservoir. This led to a slight increase in the neutral elevation of the inlet reservoir  corresponding to a zero background pressure required for the second phase. Then, in order to avoid complete depletion of the inlet reservoir by having flow in one direction only for all the experiments, the next loading phase was conducted with a flow in the opposite direction, by lowering the inlet reservoir to a negative or opposite elevation ($\Delta H_r<0$) of the exact same distance as in the upward loading phase. At the end of this first downward loading phase, which also last exactly four minutes, the inlet reservoir could simply be returned to the original zero reference elevation  having transferred back to the inlet reservoir the same amount of water that was depleted during the upward loading phase. The neutral elevations at the end of the upward and downward loading phases were found with an accuracy of $2$ to $3$ microns, which produced a very small flow below the level of detection of the $\mu$-PIV system.  This cycle was then repeated four times, with images recorded during the loading phase to measure the velocity of the backflow. 

The experimental parameters of all the pressure-relaxation experiments are presented in Table~\ref{tab:unsteadyexp}. Note that by convention the background flow mean speed $U$ is always considered positive, independently of the flow direction in the microchannel, so that the backflow is always  negative, as shown in Fig.~4 \textit{A} and \textit{B}. To indicate whether the loading phase was conducted upward or downward, $\Delta H_r$ is shown as positive or negative, respectively. As can be noticed, there is a small asymmetry of approximately $\pm 5$\% between the mean flow speed of an upward loading phase and a downward loading phase. This is due to a different spatial arrangement of the tubings between the two loading phases.
We also note that not all experiments could be exploited quantitatively. Due to the strong pressure change at the end of the loading phase, the plastron of some gratings failed and led to their wetting. Wetting could either affect the grating under study or an adjacent grating (see notes in Table~\ref{tab:unsteadyexp}). With adjacent grating failure, the backflow was still observed on the remaining plastron, with qualitatively similar magnitude and time scale, but the backflow velocity was affected through viscous stresses in the water. These data were therefore not quantitatively accurate and have not been included in the graphs presented in Fig.~4 \textit{C} and \textit{D}. The contrast between the flow field of a non-wetted grating and a wetted grating is also very clear when stopping the background pressure gradient. Similar to the flow above a ridge, the flow in the wetted grating stops immediately at the end of the loading phase, showing no backflow.

\begin{table}[tb]
    \caption{Experimental parameters of all the pressure-relaxation experiments. }
    \centering
    \label{tab:unsteadyexp}
    \begin{tabular}{c c c c}
          \toprule
          {Exp.} & {$\Delta H_r$ (\SI{}{mm})} & {$U$ (\SI{}{mm/s})} & {Note} \\
          \midrule
          {1-1} & {$70$} & {$4.4$} & {} \\
          {1-2} & {$-70$} & {$3.9$} & {} \\
          {1-3} & {$70$} & {$4.3$} & {} \\
          {1-4} & {$-70$} & {$3.9$} & {} \\
          {1-5} & {$70$} & {$4.3$} & {} \\
          {1-6} & {$-70$} & {$3.8$} & {} \\
          {1-7} & {$70$} & {$4.3$} & {} \\
          {1-8} & {$-70$} & {$3.8$} & {} \\
          {2-1} & {$100$} & {$5.4$} & {Adjacent grating failed} \\
          {2-2} & {$100$} & {$4.8$} & {Adjacent grating failed} \\
          {2-3} & {$100$} & {$5.3$} & {Adjacent grating failed} \\
          {2-4} & {$ 100$} & {$4.4$} & {Adjacent grating failed} \\
          {3-1} & {$-40$} & {$2.3$} & {} \\
          {3-2} & {$40$} & {$2.5$} & {} \\
          {3-3} & {$-40$} & {$2.3$} & {} \\
          {3-4} & {$40$} & {$2.5$} & {} \\
          {3-5} & {$-40$} & {$2.3$} & {} \\
          {3-6} & {$40$} & {$2.5$} & {} \\
          {3-7} & {$-40$} & {$2.3$} & {} \\
          {3-8} & {$40$} & {$2.4$} & {Adjacent grating failed} \\
          {3-9} & {$-40$} & {$2.2$} & {Adjacent grating failed} \\
          {3-10} & {$40$} & {$2.4$} & {Adjacent grating failed} \\
          {3-11} & {$-40$} & {$2.2$} & {Grating under study failed} \\
          {3-12} & {$40$} & {$2.5$} & {Adjacent grating failed} \\
          {4-1} & {$130$} & {$7.4$} & {Adjacent grating failed} \\
          \bottomrule
    
    \end{tabular}
\end{table}

\subsection{Impact of thermal Marangoni effects in experiments}
We assess the potential impact of thermal Marangoni effects in our experiments to examine whether the backflow observed in the pressure-relaxation experiments could be due to thermal Marangoni effects. We distinguish steady temperature gradients from flow-dependent gradients. 

Steady temperature gradients can arise due to spatial temperature variations in the setup close to the microchannel. As the backflow was observed in both flow directions through the microchannel, within a few minutes' interval, this implies that steady spatial temperature gradient did not affect the experiments.

Temperature gradients could also arise due to heat being advected by the flow during the loading phase. This heat could come from the laser, although we note that its power was already very small, less than \SI{155}{\micro\watt}, and only a small fraction would have been absorbed as heat by water. As the microchannel height is only $H\approx\SI{0.1}{\milli\meter}$, the timescale for thermal diffusion across $H$ is of the order of $H^2/D_T\approx \SI{0.1}{\second}$, where
$D_T \approx \SI{1.4e-7}{\meter\tothe{2}\per\second}$ is the thermal diffusivity of water. Since the bottom of the microchannel is maintained at a fixed temperature by a Peltier element, as soon as the loading phase ends the temperature variations in the microchannel should vanish within approximately $\SI{0.1}{\second}$. However, the time scales for the backflow are typically of the order of one minute. Therefore, thermal Marangoni effects had negligible impact in our experiments.



\section{Supporting model for pressure-relaxation experiments}
We develop a model that predicts the temporal scale for the rapid backflow that we observed in the pressure-relaxation experiments (Fig.~4). We consider the geometry  described in Fig.~2\textit{A}, with the flow in the loading phase  in the positive direction. We assume that the backflow observed at the end of the loading phase and developing in the negative direction is dominated by advection, and therefore neglect adsorption/desorption and diffusion along the interface. We solve the one-dimensional time-dependent advection equation for the transport of  surfactants at the interface.  Equation (\ref{eq:Gamma}) simplifies to
\begin{equation}\label{eq:GammaAdvection}
\frac{\partial \Gamma}{\partial t} + \frac{\partial (u_I \Gamma)}{\partial x}= 0.
\end{equation}
%

Since $Re = U H / \nu \ll 1$, viscous spreading across the channel height occurs very fast compared to surfactant advection, and we approximate $(\partial u /\partial y)_I$ as $u_I/H$ in the left-hand side of (\ref{eq:MarangoniStress}). Given that the surfactant concentration is very small compared to $\Gamma_\mathrm{m}$, the right-hand side of (\ref{eq:MarangoniStress}) can be linearized to give
\begin{equation}\label{eq:viscousMarangoni}
\mu \frac{u_I}{H}= -n R T \, \frac{\partial \Gamma}{\partial x},
\end{equation}
Substituting $u_I$ from (\ref{eq:viscousMarangoni}) into (\ref{eq:GammaAdvection}) we find the conservation equation for $\Gamma$:
\begin{equation}\label{eq:GammaJulien}
\frac{\partial \Gamma}{\partial t} -H \frac{n R T}{\mu} \frac{\partial }{\partial x} \left(\Gamma \frac{\partial \Gamma}{\partial x} \right)= 0.
\end{equation} 
We introduce similarity variables
\begin{eqnarray}
\eta &=& \frac{(g+\ell/2-x)}{g} \left( \frac{t U_\Gamma}{g} \right) ^{-1/3} , \\ 
\quad \frac{\Gamma(x,t)}{\Gamma_\mathrm{m}} &=& \left( \frac{t U_\Gamma}{g} \right)^{-1/3} f(\eta) ,
\end{eqnarray}
where $U_{\Gamma}=nRT\Gamma_\mathrm{m}/\mu$ is a Marangoni-based velocity scale. Substituting into (\ref{eq:GammaJulien}), integrating twice and using the boundary condition (\ref{eq:SurfNoFlux}), the solution is
\begin{equation}\label{eq:gammadistrib}
\frac{\Gamma(x,t)}{{\Gamma_\mathrm{m}}}= -\frac{(g+\ell/2-x)^2}{6H {U_\Gamma} t}+C \left(\frac{t U_\Gamma}{g}\right)^{-1/3},
\end{equation}
which is valid for $t$ larger than diffusion time (i.e. at very small $t$ diffusion plays a role) and for time small enough that the front of the advection is still between the two stagnation points (at $x=g+\ell/2$ and $x=\ell/2$). For the initial condition, if we assume that the loading phase had a strong positive background flow $U$ (such as in the pressure-relaxation experiments), then  surfactants have accumulated near the downstream stagnation point $x=g+\ell/2$ for $t\leq 0$. The exact  distribution of the surfactants at $t=0$ is unknown, but for large loading speed $U$, we can assume that it is very steep near $x=g+\ell/2$. The constant of integration $C$ is effectively a measure of the total amount of surfactant on the interface, which is constant at all time for insoluble surfactants. This is effectively the main unknown in our experiments. We find
\begin{equation}
C= \left(\frac{\int_{x_f}^g\Gamma/\Gamma_\mathrm{m} \,dx}{2/3\sqrt{6Hg}} \right)^{2/3},
\end{equation}
where $x_f(t)$ is the front of the surfactant, i.e. where $\Gamma$ vanishes:
\begin{equation}
x_f= g+\ell/2-\sqrt{6Hg C} \left( \frac{t U_\Gamma}{g} \right)^{1/3},
\end{equation}
which is effectively valid in our finite-length geometry until the front reaches the stagnation point at $x=\ell/2$, as the model assumes a semi-infinite lane $x\leq g+\ell/2$.

From (\ref{eq:viscousMarangoni}) and (\ref{eq:gammadistrib}), the interfacial speed due to surfactant gradients is therefore,
\begin{equation}
u_I = -\frac{g+\ell/2-x}{3t}.
\end{equation}
Thus, $u_I$ is negative, corresponding to the backflow observed in our experiments (Fig.~4 \textit{A} and \textit{B})  for $t>0$. Once the loading phase ends and the background flow stops, the surfactants travel back along the interface, driving a Marangoni backflow, to eventually redistribute uniformly along the air--water interface, $\ell/2\leq x\leq g+\ell/2$. This result also shows that the backflow velocity should decrease in time as $1/t$. We compare this scaling prediction with our experimental results in Fig.~4\textit{D}. As mentioned previously, this result is not valid at very small time, where diffusion processes are important and the initial distribution of the surfactant is unknown. Hence, it cannot inform us about the dependence of the peak backflow velocity, measured at $t\approx 0$, with $U$ (Fig.~4\textit{C}). It only informs us about the trend of the backflow at intermediate times, as shown in Fig.~4\textit{D}, until the opposite stagnation point at $x=\ell/2$ starts playing a role.

\end{document}